\documentclass[preprint,showpacs,preprintnumbers,amsmath,amssymb]{revtex4}

\usepackage{graphicx}
\usepackage{dcolumn}
\usepackage{bm}

\begin{document}

\preprint{APS/123-QED}

\title{Collision Drag  Effect on Two-Fluid Hydrodynamics of Superfluid $^3$He 
in Aerogel}

\author{M. Miura, S. Higashitani, M. Yamamoto and K. Nagai}

\affiliation{%
Faculty of Integrated Arts and Sciences, Hiroshima University, Kagamiyama 1-7-1, Higashi-Hiroshima 739-8521, Japan
}%


\date{\today}

\begin{abstract}
Sound propagation  in superfluid $^3$He in aerogel is studied 
on the basis of a two-fluid model taking into account 
the effect by the drag force due to collisions between 
$^3$He-quasiparticles and aerogel molecules. 
The drag force plays a role of frictional force between the aerogel 
and the normal-fluid component. 
In local equilibrium, they move together in accordance with 
McKenna {\it et al.}'s model. 
The deviation from the local equilibrium leads to the damping
of sound. We give explicit expressions for the attenuation of longitudinal
sounds in this system. 
We also discuss 
the sound propagation in a superfluid ${}^3$He-aerogel system 
embedded in a narrow pore. 
It is shown that the forth sound propagates in such a system because of 
the clamping of the normal fluid by the aerogel. 
\end{abstract}

\pacs{Valid PACS appear here}
\maketitle
\section{Introduction}
It is well known that there exists a variety of sound modes in liquid $^3$He. 
In the normal phase, 
in addition to the usual hydrodynamic 
sound mode (the first sound), the so-called zero sound propagates 
as a well-defined sound mode in the collisionless regime 
at low temperatures. In the superfluid phase,
other two kinds of sound, the second and forth sound, can propagate 
which reflects the presence of the normal and the superfluid 
component of liquid. 
The second sound is the out-of-phase oscillation of the two fluids. 
The forth sound is the oscillation of only 
the superfluid component when the normal fluid is clamped owing to 
its viscosity in a narrow channel. 

Recently, there has been increasing interest in 
the sounds in liquid $^3$He confined in highly 
porous aerogel.\cite{Nomura,Gervais,McKenna,Golov,Rainer,Ichikawa,Higashitani,Higa_lt23,Kotera,Berkutov} 
The aerogel consists of a random structure of silica strands 
of approximately 3 nm in diameter. The $^3$He-aerogel system has been 
regarded as a model system for studying the impurity scattering effects 
on the Fermi liquid which undergoes a transition to an unconventional 
Cooper pairing state. 
Nomura {\it et al}.\cite{Nomura} and Gervais {\it et al.}\cite{Gervais} 
have observed that the longitudinal sound attenuation in normal liquid 
$^3$He in aerogel does not show the characteristic temperature dependence 
of the Fermi liquid, i.e., the first-to-zero sound transition. 
In addition, the normal phase attenuation is of the same order 
of magnitude as the pure $^3$He case in spite of the 
presence of random scatterers. They also found the absence of the 
attenuation peak due to the order parameter collective excitation 
in the superfluid phase of $^3$He in aerogel. 
To study the attenuation theoretically, 
the homogeneous scattering model\cite{Thuneberg} can be a good starting point, 
because the sound wavelength ($\sim$ 20 $\mu$m) 
is much longer than the inter-strands 
spacing ($\sim$ 30 nm),\cite{Nomura,Gervais}
so that the detailed structure of aerogel is expected not to play 
a significant role. 
The homogeneous scattering model, however, predicts a significantly 
large attenuation\cite{Rainer,Ichikawa} 
(which is two order of magnitude larger than the observed value) 
though it can reproduce the qualitative temperature dependence.

To resolve the discrepancy between the theory and the experiment,
Ichikawa {\it et al}.\cite{Ichikawa} have studied the effect of simultaneous 
aerogel motion caused by collisions with $^3$He-quasiparticles. 
Such a collision drag effect reduces drastically the longitudinal 
sound attenuation and the experimental results can be reproduced by 
the collision drag model.\cite{Ichikawa,Higa_lt23} 
This model has recently been applied 
to the transverse sound in the $^3$He-aerogel system.\cite{Higashitani}
It was shown that the collision drag effect modifies strongly the nature 
of the transverse sound propagation; for example, 
the transverse sound in the hydrodynamic regime, which is an overdamped 
mode in pure $^3$He, can propagate over a long distance. 

In this paper, we discuss the collision drag effect in superfluid
${}^3$He-aerogel system. McKenna {\it et al}.\cite{McKenna} 
have already proposed a theory of the sound in superfluid-aerogel system.
They modified a set of two-fluid hydrodynamic equations 
to take into account the interlocking motion of the normal 
component and the aerogel. 
This theory was used by Golov {\it et al.}\cite{Golov} to extract 
the superfluid density $\rho_s$ in the $^3$He-aerogel system from 
the sound velocity measurements. 

In deriving the modified two-fluid hydrodynamic equations, 
McKenna {\it et al.}\cite{McKenna} assumed the normal fluid to 
be completely locked to the aerogel matrix, i.e., 
the system is assumed to be in local equilibrium. 
The resulting equations can determine only the sound velocity and cannot 
give any information about the attenuation. 
In this paper, we discuss the longitudinal sound attenuation 
in superfluid $^3$He in aerogel when the system deviates slightly from 
local equilibrium. 
In addition to the bulk $^3$He-aerogel system, 
we shall consider a restricted geometry such that the aerogel is 
built in a pore formed by sintered silver powders, 
as used in the fourth sound experiment\cite{Kotera}.

\section{Modification of Two-fluid Hydrodynamic Equations }
Now we consider the drag force exerted on aerogel molecules. 
The drag force can be defined as the net momentum transfer 
per unit time from superfluid ${}^3$He 
to the aerogel during impurity scattering 
processes. Thus the drag force can be determined from the collision 
integral in the quasiparticle transport 
equation.\cite{Ichikawa,Higashitani,Higa_lt23} 
In the normal state, the drag force density ${\bf F}$ 
is given by 
${\bf F} =\frac{1}{\tau_{\rm tr}}(1+F^s_1/3)\rho({\bf v}-{\bf v}_a)$,
where $\tau_{\rm tr}$ is the transport relaxation time for the impurity 
scattering, $\rho$ the mass density of liquid $^3$He, 
$F_1^s$ the Landau parameter 
and ${\bf v}$ and ${\bf v}_a$ the local velocities of the liquid 
and the aerogel, respectively. Since ${\bf F}$ is proportional to the relative 
velocity ${\bf v} - {\bf v}_a$, it can be also interpreted as 
friction between the liquid and the aerogel. In the superfluid phase, 
the motion of the superfluid component is expected to uncouple 
with the aerogel. 
It is, therefore, plausible to generalize the normal-state result of 
${\bf F}$ into 
\begin{equation}
{\bf F}=\frac{1}{\tau_{\rm f}}\rho_n({\bf v}_n-{\bf v}_a),
\label{fdef}
\end{equation}
where $\rho_n$ is the normal-fluid density, ${\bf v}_n$ is the normal-fluid 
velocity and $\tau_{\rm f}$ denotes a relaxation time 
characterizing the frictional effect between the normal fluid and the aerogel. 
The relaxation time $\tau_{\rm f}$ is expected to be of the same order
as the impurity scattering time $\tau_{\rm i}$. In fact in the normal
phase it is given by
$\tau_{\rm tr}/(1+F_1^s/3)$. In superfluid phase, involved
microscopic
calculations are necessary to obtain an explicit expression for
$\tau_{\rm f}$. 
We shall, therefore, 
treat $\tau_{\rm f}$ as a phenomenological parameter in this paper.

We can now obtain two-fluid hydrodynamic equations taking into account 
the collision drag effect. In the set of the conventional two-fluid 
equations, the momentum conservation law is modified 
because of the momentum loss due to the drag force, while  
the equation of motion of the superfluid velocity ${\bf v}_s$ remains 
unchanged. The two-fluid equations for longitudinal sound are thus given by
\begin{eqnarray}
&&\frac{\partial \rho}{\partial t}+{\rm div}\,{\bf J}=0 ,
\label{mass}\\
&&\frac{\partial{\rm div}\,{\bf J}}{\partial t} = -\nabla^2 P 
+ \frac{4}{3}\eta\nabla^2{\rm div}\,{\bf v}_n - {\rm div}\,{\bf F} , 
\label{momentumcon}\\
&&\frac{\partial {\bf v}_s}{\partial t}=-\frac{1}{\rho}{\nabla}P+S{\nabla}T ,
\label{supereq}\\
&&\frac{\partial{\rho}S}{\partial t}+{\rm div}\,({\rho}S{\bf v}_n)=0 ,
\label{entropy}
\end{eqnarray}
where $\rho = \rho_n + \rho_s$, 
${\bf J} = \rho_n{\bf v}_n+\rho_s{\bf v}_s$, 
$\eta$ is the viscosity 
and $P$, $S$ and $T$ are pressure, entropy and temperature, respectively. 
The above equations are different from 
those by McKenna {\it et al}.\cite{McKenna} in that 
the drag force ${\bf F}$ and the viscosity $\eta$ are included. 
The former enables us to study the case when ${\bf v}_n \neq {\bf v}_a$. 
The latter is necessary to reproduce the normal-state dispersion 
relation derived previously.\cite{Ichikawa,Higa_lt23} 
The two-fluid equations must be supplemented 
by the equations of motion of aerogel:
\begin{eqnarray}
&&\rho_a\frac{\partial {\bf v}_a}{\partial t}=-{\nabla}P_a+{\bf F} ,
\label{aeroeq}\\
&&\frac{\partial \rho_a}{\partial t}+{\rm div}\,(\rho_a{\bf v}_a)=0 ,
\label{aeromass}
\end{eqnarray}
where $\rho_a$ is the aerogel mass density and 
the restoring force due to the elasticity of the aerogel is 
denoted as $-\nabla P_a$. The pressure $P_a$ is related to the 
longitudinal sound velocity 
$C_{al}$ of the skeleton aerogel by\cite{McKenna} 
\begin{equation}
C_{al}^2 = \partial P_a/ \partial\rho_a .
\label{cadef}
\end{equation}
\section{Sound Dispersion Relation}
From the above set of equations, we can derive the dispersion 
relation of the longitudinal sound in superfluid $^3$He-aerogel system. 
To do that, we shall assume, as usual, that all the non-equilibrium 
quantities have the space and time dependence of 
$\exp(i{\bf q}\cdot{\bf r} - i\omega t)$. 
Then, using eq.\ (\ref{fdef}) 
and the equations of motion of aerogel, we obtain the 
following relation between ${\bf v}_n$ and ${\bf v}_a$: 
\begin{equation}
(1-\frac{\omega_q^2}{\omega^2})\rho_a{\bf v}_a =
\frac{i}{\omega\tau_{\rm f}}\rho_n({\bf v}_n - {\bf v}_a)
\label{relav}
\end{equation}
where $\omega_q = C_{al}q$ is the frequency of skeleton aerogel sound. 
It follows that the local equilibrium (${\bf v}_n = {\bf v}_a$) is 
achieved by taking the limit $\omega\tau_{\rm f} \rightarrow 0$. 
Combining eq.\ (\ref{relav}) with our two-fluid equations, 
we find that the dispersion relation can be written as 
\begin{equation}
(z^2-C_1^2)(z^2-C_2^2)  +(i\frac{4{\eta}{\omega}}{3{\rho_n}}
+\frac{i}{\omega\tau_{\rm f}}\frac{z^2}{1+
\frac{i}{\omega\tau_{\rm f}}\frac{\rho_n}{\rho_a}\chi_a})(z^2-C_4^2)=0
\label{dispersion}
\end{equation}
with $z=\omega/q$ and 
\begin{equation}
\chi_a = \frac{\omega^2}{\omega^2-\omega_q^2} = \frac{z^2}{z^2-C_{al}^2}.
\end{equation}
In eq.~(\ref{dispersion}), $C_1$, $C_2$ and 
$C_4=[({\rho_s}/{\rho})C_1^2+({\rho_n}/{\rho})C_2^2]^{1/2}$ 
are the first, second and fourth sound velocities 
as defined in the conventional manner. 
It is easy to show that eq.\ (\ref{dispersion}) is reduced to 
the previous result\cite{Ichikawa,Higa_lt23} 
in the normal state. Moreover, eq.\ (\ref{dispersion}) 
with $\eta = 0$ and in the limit 
$\omega\tau_{\rm f} \rightarrow 0$ coincides, as expected, 
with the dispersion relation of McKenna {\it et al}.\cite{McKenna}
\begin{equation}
(z^2-C_1^2)(z^2-C_2^2)+\frac{\rho_a}{\rho_n}(z^2-C_{al}^2)(z^2-C_4^2)=0.
\label{mck}
\end{equation}

Equation (\ref{mck}) determines the velocities of two sound modes 
(fast and slow modes) in local equilibrium.\cite{McKenna,Golov}
The two sound modes in the $^3$He-aerogel system have been observed 
by Golov {\it et al.}\cite{Golov} 
According to this experiment, the velocities of the fast and slow 
modes, $C_f$ and $C_s$, satisfy the conditions such that 
$C_f^2 \gg C_2^2$, $C_f^2 \gg C_{al}^2$ and $C_1^2 \gg C_s^2 \gg C_2^2$. 
Taking into these conditions, we find from eq.\ (\ref{mck}) that 
$C_f$ and $C_s$ are %
approximately given by
\begin{eqnarray}
&&C_f^2 = \frac{C_1^2+(\rho_a/\rho_n)C_4^2}{1 + \rho_a/\rho_n} ,
\label{fast}\\
&&C_s^2 = \frac{C_{al}^2}{1+\rho\rho_n/\rho_a\rho_s} .
\label{slow}
\end{eqnarray}

When $\tau_{\rm f}$ and $\eta$ are finite, 
the complex velocities $z$'s have corrections which give 
the attenuation $\alpha = {\rm Im}\,q$ of the two hydrodynamic sound modes. 
Using eq.\ (\ref{dispersion}) and keeping only the leading order 
correction terms, we obtain the attenuations of the fast and slow modes, 
$\alpha_f$ and $\alpha_s$, as 
\begin{eqnarray}
&&\alpha_f = \frac{\omega}{2C_f^3}\frac{C_1^2-C_4^2}{(1+\rho_a/\rho_n)^2}
(\frac{4\eta\omega}{3\rho_nC_f^2} 
+ \frac{\rho_a^2}{\rho_n^2}\omega\tau_{\rm f}) 
\label{fastatten}
\\
&&\alpha_s = \frac{\omega}{2C_s}\frac{\rho_n/\rho_a}
{1+\rho\rho_n/\rho_a\rho_s}
(\frac{4\eta\omega}{3\rho_nC_s^2} 
+ \frac{\rho^2}{\rho_s^2}\omega\tau_{\rm f}) 
\label{slowatten}
\end{eqnarray}
The expressions for the attenuations include terms $\propto \eta$ 
and $\propto \tau_{\rm f}$, reflecting the viscous effect 
and the frictional effect, respectively. The information on 
the phenomenological parameters $\eta$ and $\tau_{\rm f}$ can thus be obtained 
from attenuation measurements. We shall show in the next section that 
a more useful and direct information on the frictional 
relaxation time $\tau_{\rm f}$ can be provided by the forth sound 
attenuation measurement.
\section{Fourth Sound in Narrow Pore}
Now we discuss the sound propagation in a superfluid ${}^3$He-aerogel 
system built in a narrow pore.
Such a system has recently been prepared 
by Kotera {\it et al}.\cite{Kotera} using sintered silver powders of 
$\sim$ 70 $\mu$m in diameter to study the fourth sound propagation 
in aerogel. The sintered silver powders 
provide pores with the average diameter of $L \sim 10$ $\mu$m. 

In such a system, the sintered silver will play a role to 
clamp the aerogel strand ends. 
The longitudinal wave of the aerogel along the pore channel is, therefore,
accompanied by a transverse standing wave so that the 
fixed end boundary condition should be satisfied.
It follows that 
the eigenfrequency $\omega_q$ of the aerogel is replaced by
\begin{equation}
\omega_q^2 = C_{al}^2q^2+\left( \frac{\pi C_{at}}{L} \right)^2 ,
\end{equation}
where $C_{at}$ is the transverse sound velocity of aerogel. 
The second term gives an energy gap due to the standing wave motion. 
Using $C_{at}=178.5$ m/s for 91 \% porous aerogel\cite{Berkutov}, 
the gap at $q=0$ is estimated to be 56 MHz for $L = 10$ $\mu$m.  
We see that a condition $\omega_q^2 \gg \omega^2$ is fully satisfied 
in low-frequency acoustic experiments in which the frequencies 
are typically about 10 kHz or less. 

This observation shows that the aerogel motion in the pore 
is hardly excited by the low frequency sounds. 
In the limit $\omega^2/\omega_q^2 \rightarrow 0$, 
the aerogel cannot move and then the dispersion relation 
(\ref{dispersion}) takes the form
\begin{equation}
(z^2-C_1^2)(z^2-C_2^2)  +(i\frac{4{\eta}{\omega}}{3{\rho_n}}
+\frac{i}{\omega\tau_{\rm f}}z^2)(z^2-C_4^2)=0 .
\label{disppore}
\end{equation}
The fourth sound can propagate in the limit 
$\omega\tau_{\rm f} \rightarrow 0$ because the normal fluid is clamped 
by the ``hard'' aerogel. 
The attenuation $\alpha_4$ of the fourth sound is obtained from 
eq.\ (\ref{disppore}). The result is given, up to leading order 
in $\omega\tau_{\rm f}$, by
\begin{equation}
\alpha_4
=\frac{\omega}{2C_4^3}(C_1^2-C_4^2)\omega\tau_{\rm f} .
\label{forthatten}
\end{equation}
Note that eq.\ (\ref{forthatten}) does not involve the viscosity $\eta$ 
because its contribution to $\alpha_4$ 
occurs only in higher order terms in the $\omega\tau_{\rm f}$ expansion. 
The attenuation of the fourth sound in the $^3$He-aerogel system 
is thus dominated by the frictional relaxation time $\tau_{\rm f}$. 
Using eq.\ (\ref{forthatten}), the parameter $\tau_{\rm f}$ can be 
determined from simultaneous measurements of $C_4$ and $\alpha_4$. 

The reason why the frictional effect dominates the fourth sound damping 
is the following. In the pore channel with the aerogel,
the normal flow no longer obeys the Hagen-Poiseuille law but is
governed by the Drude law.\cite{Einzel} In the presence of aerogel,
the viscous penetration depth $\delta$ for $\omega\tau_{\rm f} \ll 1$ is given
by\cite{Einzel}
\begin{equation}
 \delta =\sqrt{\frac{\eta\tau_{\rm f}}{\rho_n}}
\end{equation}
and the viscosity in the superfluid phase is
\begin{equation}
 \eta\sim \rho_n {\bar v}_{\rm F}^2 \tau_{\eta},
     \end{equation}
where ${\bar v}_F$ is an appropriate average of the quasi-particle group velocity.
At low temperatures, the viscous relaxation time $\tau_{\eta}$ is also
limited by the impurity scattering time $\tau_{\rm i}\sim \tau_{\rm f}$.
In fact, in case of the sound experiment by Nomura { \it et
al}.\cite{Nomura}
(98\% aerogel, 16 bar) one can see that $\tau_{\eta}\sim \tau_{\rm i}$
happens at 10 mK. We find therefore that the viscous penetration depth
in the presence of the aerogel is given by
\begin{equation}
 \delta\sim {\bar v}_{\rm F}\tau_{\rm f}
\end{equation}
and is as small as the mean free path.
\section{Concluding Remarks}
We have discussed hydrodynamic sound propagation in
superfluid ${}^3$He-aerogel system. In a bulk system,
the aerogel  oscillates together with the normal liquid by the collision drag
effect. As a result, two kinds of sound (the fast mode and the slow
mode)\cite{McKenna} can propagate as was observed by Golov 
{\it et al.}\cite{Golov}. The deviation in the local velocity between
the normal component and the aerogel leads to the damping of the sounds.
We have given explicit expressions for the sound absorption.
On the contrary in a narrow pore, the aerogel cannot move because
the aerogel is clamped by the pore surface. In that case, the
fourth sound can propagate along the pore channel.  
It is to be emphasized that the inter-locking between the normal
component and the aerogel occurs not by the viscous effect as so far
considered but by the collision drag effect. In other words, the
normal component is in the Drude regime.\cite{Einzel}
 
The modified two-fluid theory presented in this paper is
still a phenomenological theory and includes many unknown
parameters. 
In addition to the superfluid mass density, microscopic
information on the relaxation times in dirty superfluid ${}^3$He
is necessary for more detailed comparison with experiments. 
This problem and also the justification of our two-fluid model 
shall be discussed elsewhere. 

\acknowledgments

This work is supported by a Grant-in-Aid for COE Research 
(No. 13CE2002) from the Ministry of Education, Culture, Sports, Science and 
Technology of Japan.

\end{document}